\documentclass[a4paper,12pt]{article}

\usepackage{geometry,slashed}

\usepackage
{
   amsmath,amssymb,graphicx,tabularx
}
\usepackage{soul}
\usepackage[usenames,dvipsnames]{xcolor} 
\usepackage{color}
\usepackage[verbose,colorlinks=true,naturalnames=true,linkcolor=black,urlcolor=black, citecolor=black]{hyperref}
\usepackage{caption,subcaption}
\usepackage{authblk}
\usepackage{epstopdf}
\usepackage{enumerate}
\usepackage{multirow}
\usepackage{siunitx}
\usepackage{float}
\usepackage{booktabs}
\usepackage{cite}
\usepackage{pdflscape} 
\usepackage{listings}
\usepackage{diagbox} 
\DeclareMathAlphabet{\pazocal}{OMS}{zplm}{m}{n}
\graphicspath{ {./}}
\usepackage{url} 
\usepackage{varwidth}
\usepackage{comment}
\usepackage[parfill]{parskip}

\usepackage{lineno}

\newcommand{\madgraph}{\textsc{MadGraph5\_aMC@NLO}}

\newcommand{\pythia}{\textsc{Pythia8}}

\newcommand{\ttbb}{$t\bar{t}b\bar{b}$}

\numberwithin{equation}{section}

\title{Five-flavour scheme predictions for $t\bar{t}b\bar{b}$ at next-to-leading order accuracy}

\author{Rikkert Frederix$^1$\thanks{rikkert.frederix@fysik.lu.se} }
\author{Tetiana Moskalets$^{2,3}$\thanks{tetiana.moskalets@cern.ch} }

\affil{\small $^1$Division of Particle and Nuclear Physics, Department of Physics, Lund University, Box 118, SE-221 00 Lund, Sweden}
\affil{\small $^2$Institut of Physics, Freiburg University, Hermann-Herder Str. 3a, 79104 Freiburg, Germany}
\affil{\small $^3$Department of Physics, Southern Methodist University, PO Box 750399 Dallas TX 75275, United States of America}

\begin{document}

\maketitle

\begin{abstract}
We compute top quark pair production in association with a bottom
quark pair at the LHC within the five-flavour scheme, matched to a
parton shower, employing the FxFx merging scheme for
$t\bar{t}+\textrm{jets}$ production with up to 2 jets at NLO
accuracy. To enhance the selection efficiency for the events with
$b$-jets within the inclusive five-flavour sample, we augment the
generation probability of bottom quark flavours in the short-distance
event generation. Our analysis reveals some differences from NLO
predictions within the four-flavour scheme.
\end{abstract}

\section{Introduction}

Top quark pair production in association with a bottom quark pair
(\ttbb) is pivotal for probing the fundamental interactions of the
Standard Model and extracting crucial information about its
properties~\cite{CMS:2020grm,CMS:2019eih,ATLAS:2018fwl,CMS:2023xjh}. At
the Large Hadron Collider (LHC), \ttbb\ production serves as a
significant background process across various high-energy physics
phenomena, profoundly impacting the precise determination of the top
quark Yukawa coupling from experimental data. Specifically,
\ttbb\ production represents a notable background in scenarios
involving the associated production of the Higgs boson with a top
quark
pair~\cite{ATLAS:2021qou,ATLAS:2023cbt,ATLAS:2017fak,CMS:2018hnq,CMS:2018uxb},
followed by the Higgs decay into bottom quarks, as well as in the
production of four-top quark final
states~\cite{ATLAS:2023ajo,CMS:2023ftu}.

In the simulation of the \ttbb\ process, there are two primary
theoretical frameworks: the four-flavour scheme (4FS) and the
five-flavour scheme (5FS). The 4FS approach treats bottom quarks as
massive particles, decoupled from the Parton Density Functions (PDFs)
and renormalised on-shell. While, in principle straightforward to
apply at fixed
order~\cite{Bredenstein:2008zb,Bredenstein:2009aj,Bevilacqua:2009zn,Bredenstein:2010rs,Buccioni:2019plc,Denner:2020orv,Bevilacqua:2021cit,Bevilacqua:2022twl},
challenges arise due to the large mass difference between the top and
bottom quarks. Indeed, large logarithms can appear in perturbative
calculations for multi-scale processes. This is corroborated by
difficulties in choosing optimal renormalisation and factorisation
scales. Moreover, when integrating this scheme with parton shower
simulations~\cite{Cascioli:2013era,Jezo:2018yaf} radiation generated
by the parton shower can produce additional bottom quarks. It is
currently poorly understood how this radiation should be constraint
such that the leading bottom quarks in the events are attended for by
the short-distance hard matrix element calculations, and only the
subleading bottom quarks are of parton-shower origin.

Alternatively, the 5FS accommodates massless bottom quarks within the
PDFs and adopts the $\overline{\text{MS}}$ renormalisation scheme. In
this scheme one has to generate an inclusive $t\bar{t}+\textrm{jets}$
sample~\cite{Frixione:2003ei,Frixione:2007nw,Hoeche:2014qda,Mazzitelli:2021mmm},
and select $b$-jets only after parton showering. Because the bottom
quarks are treated as massless partons, no large logarithms due to
small quark masses appear in the short-distance calculation. Moreover,
potential large scale hierarchies between the top quarks and the jets
can effectively be resummed and taken into account by a multi-jet
merging procedure. In addition, in multi-jet merging approaches the
jets generated by the parton shower are always softer than a ``merging
scale'' (except for jets coming from the highest-multiplicity sample),
which in itself is smaller than the softest jets generated by the
matrix elements, resulting in an accurate parton-shower approximation
for all jets. This is not necessarily the case for additional jets in
the 4FS approach. In practice, the parton shower
simulations~\cite{Bierlich:2022pfr,Bellm:2015jjp,Bothmann:2019yzt}
incorporate non-zero bottom quark mass effects into their splitting
functions, ensuring reasonable modeling accuracy, especially in the
infrared (IR) regions, where resummation of logarithms is particularly
important. Consequently, the 5FS approach offers a superior
description of the \ttbb\ process at the LHC as compared to the 4FS.

However, implementing the 5FS approach can be computationally
demanding, particularly when event generation involves multi-jet
merging at next-to-leading order (NLO) accuracy. Indeed, generating
$t\bar{t}+\textrm{jets}$ events with up to 2 jets at NLO
accuracy~\cite{Hoeche:2012yf,Frederix:2012ps,Platzer:2012bs}
requires substantional computing resources.  Moreover, the requirement
to select $b$-jets post-parton showering results in extremely low
event selection efficiencies, as a significant portion of generated
events do not contain additional $b$-jets, leading to an enormous
computational overhead. To address these challenges, this study
proposes a novel method to enhance the $b$-jet selection efficiency in
the 5FS approach, making it a competitive alternative to the 4FS
framework.

An alternative approach, known as the ``fusion''
method~\cite{Hoche:2019ncc,Ferencz:2024pay}, merges aspects of both
the 4FS and 5FS calculations. This method can be seen as correcting a
5FS calculation by including mass effects directly in the
short-distance calculation. Also this method would benefit from the
increased efficiency in selecting events with $b$-jets in the 5FS
sample. Note that in Ref.~\cite{Ferencz:2024pay} the 5FS component was
only computed at LO accuracy (apart from the
$t\bar{t}+0\,\textrm{jets}$ contribution), and therefore the selection
efficiency is of less importance in reducing overall CPU costs.

The outline of this paper proceeds as follows: in the next section, we
detail our proposed method for enhancing $b$-jet selection efficiency
within the 5FS simulation. Section \ref{result} presents predictions
for the \ttbb\ process in the 5FS at the 13~TeV LHC and compares them
to the 4FS. Finally, in Section \ref{conclusion}, we summarise our
findings.

\section{Enhancing $b$ flavour in Matrix Elements and Parton Showers}\label{enhance}

Predicting $b$-jet associated top quark pair production in the
5FS at the next-to-leading order plus parton
shower (NLO+PS) level presents challenges due to the dual sources of
bottom quark production: the short-distance contribution from matrix
elements and the parton shower. This duality poses efficiency issues,
where a majority of events lack bottom quarks, resulting in enormous
inefficiencies during event selection.

In order to improve the efficiency of generating bottom quarks at the
short-distance matrix element level the following improvements can be
made. During phase-space integration and unweighting, for each
contribution to an event that contains external bottom quarks, the
weight is multiplied by $w_{\mathrm{enh.}}$, irrespective if these
bottom quarks are initial or final state. This skews the event
generation to favor events with external bottom quarks. In order to
compensate for this, events selected with external bottom quarks
have their weight multiplied by $1/w_{\mathrm{enh.}}$, while the
original weight is retained for other events. This increases the
number of events with bottom quarks, while not changing any of the
physics when the event weights are taken into account.

We have implemented this procedure in the
\madgraph~\cite{Alwall:2014hca} code. The enhancement factor
$w_{\mathrm{enh.}}$ can be set by a new parameter,
\verb|bflav_enhancement|, in the \verb|run_card.dat| file. This new
feature will become part of an upcoming release of the
\madgraph\ framework. In the mean time, a version of \madgraph\ with
this feature incorporated is available from the authors upon request.

A similar biasing strategy can be applied in the parton showering
process. In versions starting from 8.311,
\pythia~\cite{Bierlich:2022pfr} offers a built-in mechanism for
enhancing splitting probabilities for certain types of
particles. While this can be used to increase the production of bottom
quarks, in practice we have found significant trade-offs. Even with a
modest increase of the $g\to b\bar{b}$ splitting probability the
weights of the events varies widely, significantly hampering the
statistical significance of the event sample, and nullifying the
improvements in the $b$-jet selection efficiency.

Alternatively, considering the faster speed of event showering
compared to short-distance event generation, a more effective approach
involves showering events without short-distance bottom quarks
multiple times ($N_{\mathrm{PS}}$) with different random number seeds
and subsequently scaling the weights of these events by
$1/N_{\mathrm{PS}}$. This approach mitigates the issue of large
weights in short-distance events without bottom quarks while
increasing the number of events containing at least one $b$-jet. Care
must be taken to set $N_{\mathrm{PS}}$ judiciously to avoid too large
correlations among short-distance events passing the selection
procedure multiple times for different shower generations.

\section{Results}\label{result}
The 5FS calculation for \ttbb\ production at the 13~TeV LHC has been
performed using the FxFx merging scheme~\cite{Frederix:2012ps} as
implemented in the \madgraph\ event
generator~\cite{Alwall:2014hca}. This involves merging
$t\bar{t}+\textrm{jets}$ processes with up to 2 jets at NLO accuracy
and matching them to the \pythia\ parton
shower~\cite{Bierlich:2022pfr}. We set the bottom-flavour enhancement
factor equal to $w_{\mathrm{enh.}}=100$ to increase the number of
short-distance events with bottom quarks\footnote{This enhancement
factor results in an expected $\times5.5$ increase in the number of
events in the two-$b$-jet scenario.}. Central values for the
renormalisation and factorisation scales are taken from the FxFx
merging procedure, with a merging scale of 40~GeV (with 70 and 100~GeV
as alternatives) and a default shower starting scale based on $H_T/2$,
with an alternative value based on $H_T/4$. Uncertainties around the
central value are estimated by considering the envelope of predictions
based on the usual 7-point scale variation, merging scale
alternatives, and shower starting scale alternative. In practice, for
most observables, the uncertainty is dominated by the renormalisation
and factorisation scale dependence.

The 5FS predictions are compared to an NLO+PS prediction in the 4FS
using the MC@NLO matching~\cite{Frixione:2002ik} as implemented in the
\madgraph\ event generator. Central values of the renormalisation and
factorisation scales are determined as follows:
\[
\mu_R=\big(E_{T,t}E_{T,\bar{t}}E_{T,b}E_{T,\bar{b}}\big)^{1/4} \qquad
\mu_F=\tfrac{1}{2}\big(E_{T,t}+E_{T,\bar{t}}+E_{T,b}+E_{T,\bar{b}}\big),
\]
respectively, with $E_T=\sqrt{m^2+p_T^2}$, following
Ref.~\cite{LHCHiggsCrossSectionWorkingGroup:2016ypw}. For the 4FS we
show the scale dependence by performing a 7-point variation for the
renormalisation and factorisation scales. Note that we do not include
shower scale uncertainties, and no matching scheme uncertainties. In
particular the latter are expected to be sizeable based on the
differences found among the various approaches considered in
Ref.~\cite{LHCHiggsCrossSectionWorkingGroup:2016ypw}, but are
non-trivial to assess exactly.

Both 5FS and 4FS events are matched to the \pythia\ parton shower. To
simplify the analysis and focus on differences between the 4FS and 5FS
setups, we do not include hadronisation, underlying events, and keep
the top quarks stable. Events that do not contain short-distance
bottom quarks have been showered $N_{\mathrm{PS}}=10$ times.

Jets are reconstructed from all final-state partons (excluding the top
quarks) using the anti-$k_T$
algorithm~\cite{Cacciari:2008gp,Cacciari:2011ma} with $\Delta
R>0.4$. Jets must have a transverse momentum $p_T>25~\textrm{GeV}$ and
pseudo-rapidity $|\eta|<2.5$. Jets containing at least one bottom
quark are identified as $b$-jets\footnote{No flavour-aware clustering
algorithm (as first discussed in~\cite{Banfi:2006hf}) needs to be
applied. In the MC@NLO matching prescription the NLO divergences are
subtracted using a massless 5FS NLO expansion of the parton
shower. The shower then re-instates the subtracted contributions
including the bottom quark mass effects.}. We consider two scenarios:
the one $b$-jet setup, where at least one $b$-jet is required for an
event to pass the selection cuts, and the two-$b$-jet setup, where at
least two $b$-jets are required.

\begin{figure}[!t]
\centering
\includegraphics[width=0.45\textwidth]{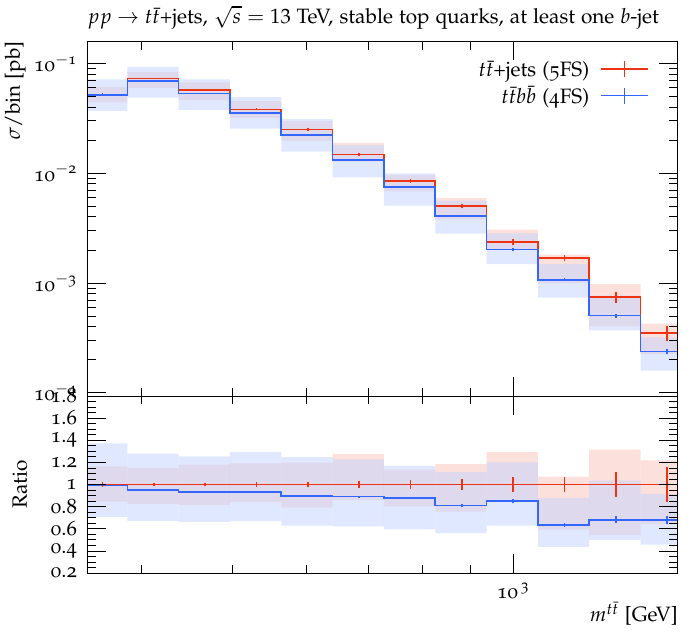}
\includegraphics[width=0.45\textwidth]{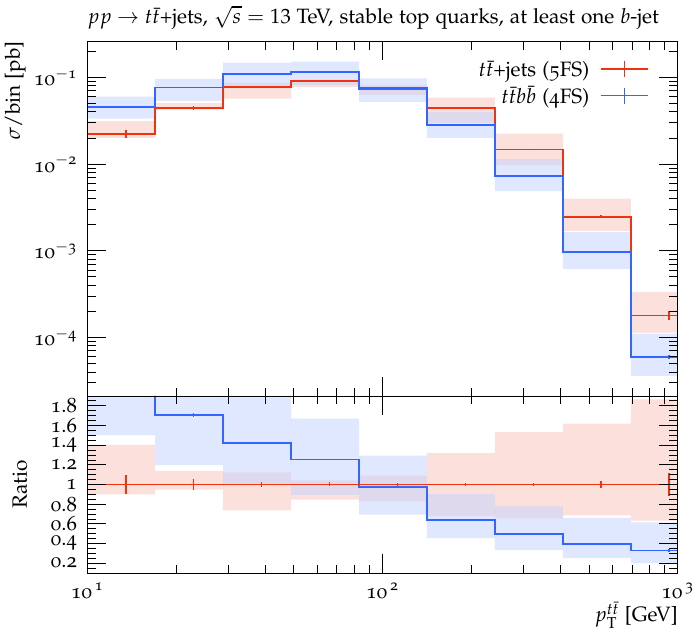}
\includegraphics[width=0.45\textwidth]{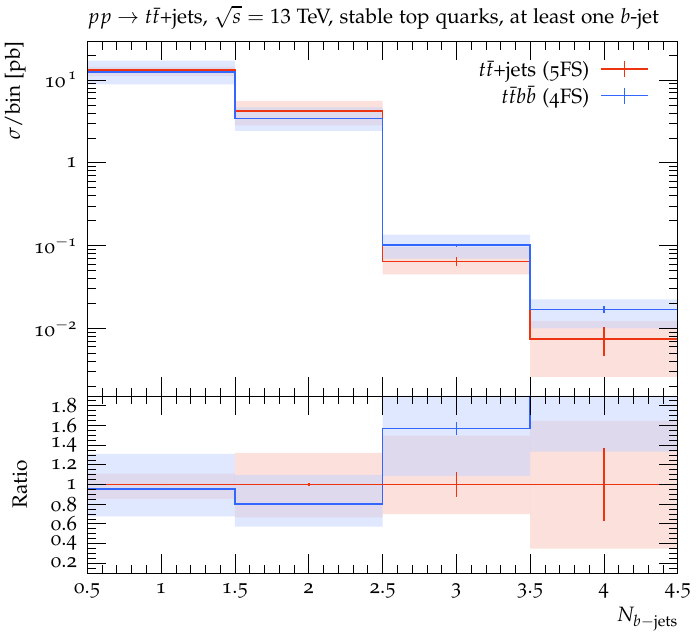}
\includegraphics[width=0.45\textwidth]{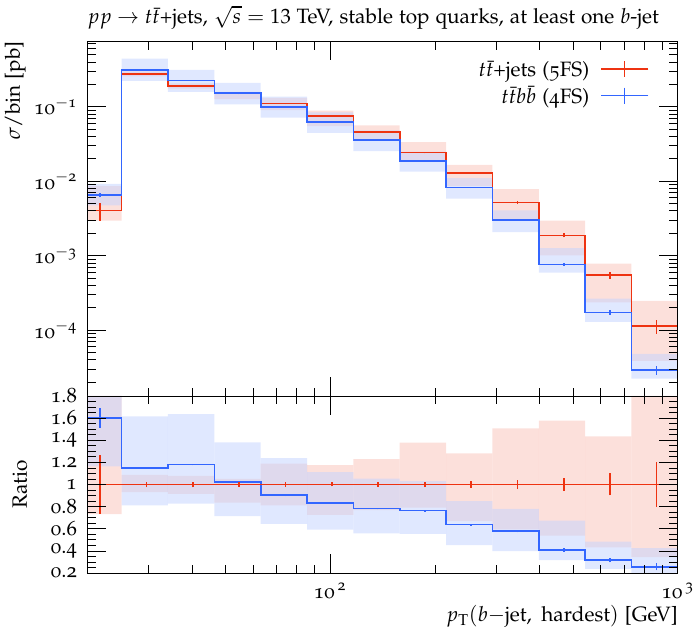}
\includegraphics[width=0.45\textwidth]{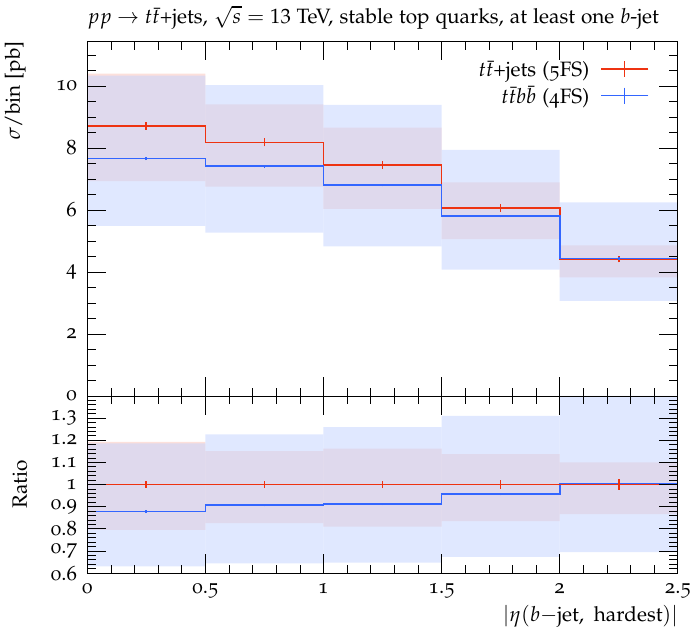}
\includegraphics[width=0.45\textwidth]{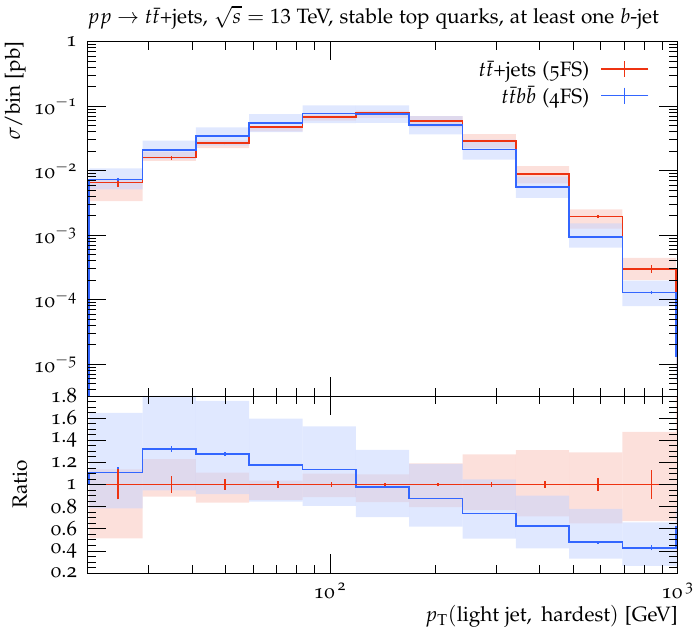}
\caption{Predictions in the 5FS and 4FS in the one-$b$-jet scenario for
  invariant mass of the $t\bar{t}$ pair (top left); Transverse
  momentum of the $t\bar{t}$ pair (top right); number of $b$-jets
  (middle left); transverse momentum of the hardest $b$-jet (middle
  right); and pseudo-rapidity of the hardest $b$-jet (bottom left);
  the transverse momentum of the hardest light jet (bottom right).}
\label{fig_1b}
\end{figure}

\begin{figure}[!t]
\centering
\includegraphics[width=0.45\textwidth]{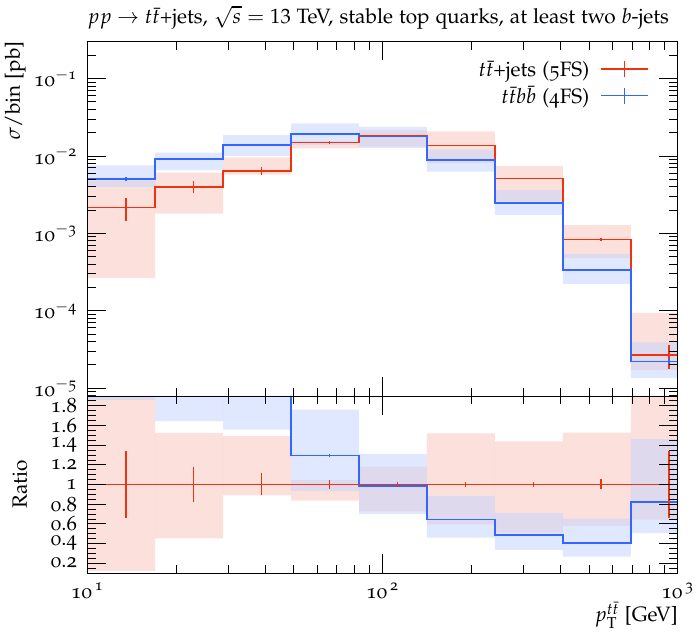}
\includegraphics[width=0.45\textwidth]{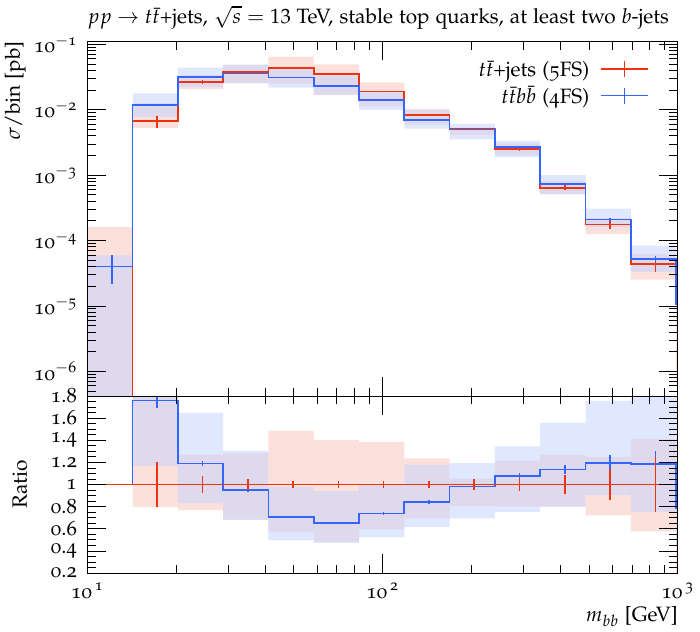}
\includegraphics[width=0.45\textwidth]{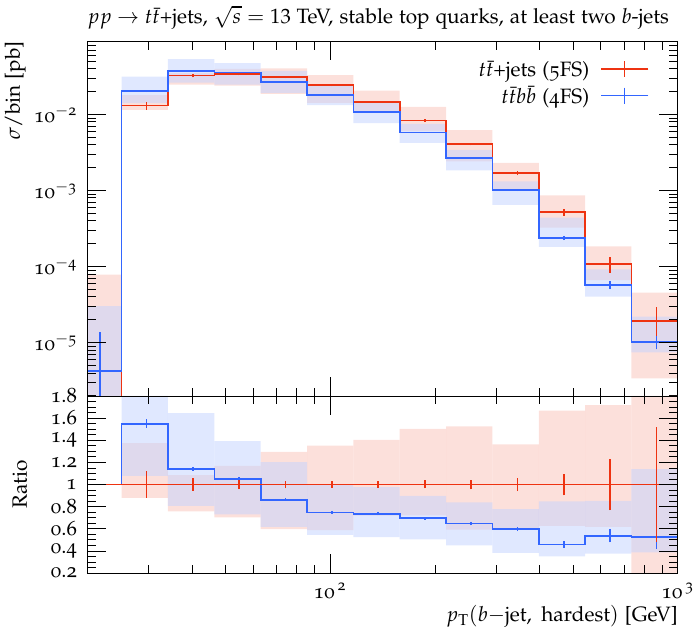}
\includegraphics[width=0.45\textwidth]{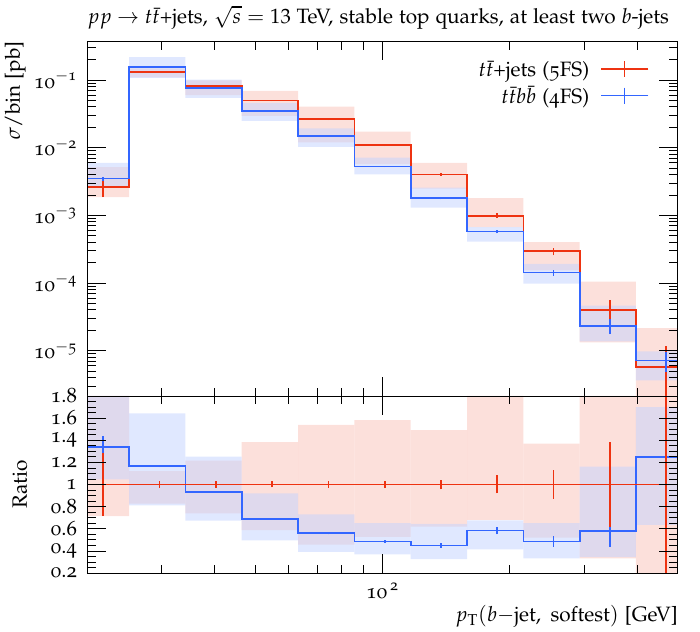}
\includegraphics[width=0.45\textwidth]{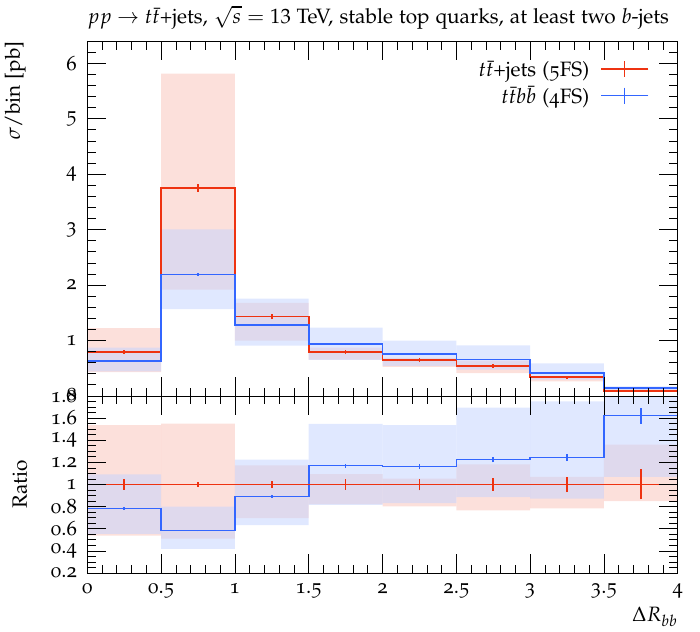}
\includegraphics[width=0.45\textwidth]{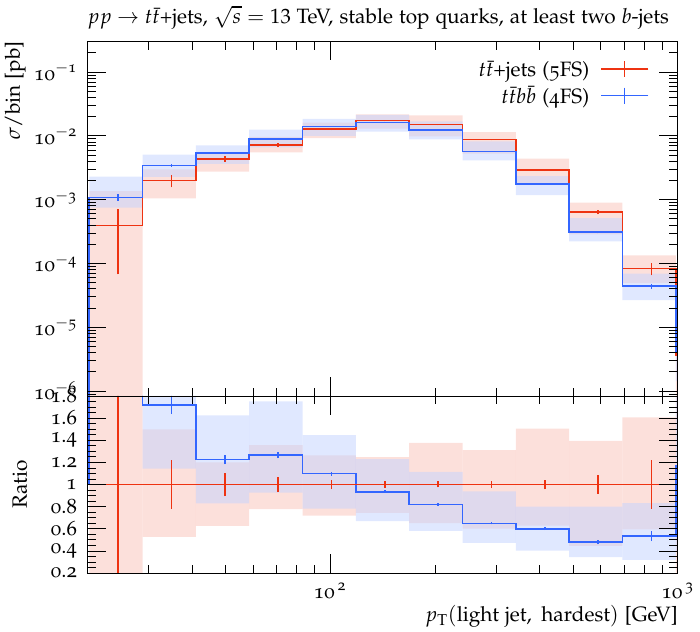}
\caption{Predictions in the 5FS and 4FS in the two-$b$-jet scenario for
  transverse momentum of the $t\bar{t}$ pair (top left); invariant
  mass of the two leading $b$-jets (top right); transverse momentum of
  the hardest $b$-jet (middle left); transverse momentum of the
  softest $b$-jet (middle right); $\Delta R$ separation between
  the two leading $b$-jets (bottom left); and the transverse momentum
  of the hardest light jet (bottom right).}
\label{fig_2b}
\end{figure}

In Figs.~\ref{fig_1b} and \ref{fig_2b}, we present a selection of
representative predictions in the one-$b$-jet and two-$b$-jet
selection scenarios, respectively. All plots follow a similar format,
created with the Rivet analysis toolkit~\cite{Buckley:2010ar}. The
main panel displays the absolute predictions, i.e., cross section per
bin, in the 5FS and 4FS schemes using red and blue histograms,
respectively. For the 5FS, the coloured band represents the envelope
of the renormalisation, factorisation, merging, and shower starting
scale variation. In the lower panel, the ratio with respect to the
central value in the 5FS is shown.

We commence with the results for the one-$b$-jet scenario. In the
upper-left plot of Fig.~\ref{fig_1b}, the top quark pair invariant
mass is depicted. The difference between the 5FS and the 4FS is
minimal, reaching approximately 20\% around $m^{t\bar{t}}\gtrsim
1~\textrm{TeV}$, with the central value of the 5FS spectrum being
harder than the 4FS. This discrepancy falls within the 5FS scale
uncertainty band. Conversely, the transverse momentum of the top quark
pair exhibits a difference (upper-right plot), with the 5FS prediction
notably harder than the 4FS one. At small transverse momenta
$p_T^{t\bar{t}}\lesssim 100~\textrm{GeV} $, the 4FS yields a larger
prediction, exceeding a factor of 2 at $p_T^{t\bar{t}}\lesssim
15~\textrm{GeV}$. Conversely, at large transverse momenta
$p_T^{t\bar{t}}\gtrsim 300~\textrm{GeV} $, the 5FS predicts more than
twice as many events as the 4FS calculation. Although the uncertainty
band of the 5FS widens at large transverse momenta, the central value
of 4FS lies outside of this band across almost the entire range
considered. When taking the renormalisation and factorisation scale
dependence of the 4FS into account, the bands almost touch over the
range considered. Given that the true 4FS uncertainty is expected to
be larger than just the renormalisation and factorisation scale
dependence, one can conclude that the two schemes are in agrement for
this observable. However, the difference in shape between the two
central values is rather notable, and in appendix A we further
investigate this observable.

Turning our attention to the $b$-jets (not stemming from the decay of
the top quarks, as the top quarks are kept stable in the predictions),
we depict the number of $b$-jets in the middle-left plot, the
transverse momentum of the hardest $b$-jet (i.e., the one with the
highest transverse momentum) in the middle-right plot, and the
pseudo-rapidity of the hardest $b$-jet in the lower-left plot of
Fig.~\ref{fig_1b}. The cross section for exactly 1 and 2 $b$-jets
exhibits similarity between the 5FS and 4FS calculations, with their
respective central values well within the uncertainty band of the
other. Conversely, for 3 or 4 $b$-jets, the cross section predicted
in the 4FS is somewhat larger than in the 5FS, although uncertainties
are significant for these bins. The central value of the 4FS
prediction for the transverse momentum of the hardest $b$-jet lies on
the edge of the uncertainty band associated with the 5FS prediction,
with the 5FS yielding a harder spectrum. Regarding the pseudo-rapidity
of this jet, the two predictions are in good agreement.

The transverse momentum distribution of the hardest light jet (i.e.,
the jet with the highest transverse momentum not containing any bottom
quarks) is presented in the lower-right plot of Fig.~\ref{fig_1b}. As
observed from this distribution, the 5FS prediction results in a
harder spectrum for this jet, with the central value of the 4FS
spectrum lying outside of the 5FS uncertainty band, but the two bands
do show some overlap over the entire range considered.

Similarly to the one-$b$-jet scenario, also in the two-$b$-jet
scenario, Fig.~\ref{fig_2b} shows that the transverse momentum of the
top quark pair is significantly harder in the 5FS compared to the 4FS
scheme (top-left plot). Again, the central value of 4FS curve lies
outside of the 5FS uncertainty band, with the bands themselves barely
touching, see also appendix A. For the invariant mass between the two
leading $b$ jets, i.e., the two $b$-jets with the highest transverse
momenta, the 5FS and 4FS predictions are in agreement, as seen in the
upper-right plot of Fig.~\ref{fig_2b}. The 4FS prediction lies within
the 5FS uncertainty band. Both the transverse momentum distributions
of the hardest and the softest $b$-jet, shown in the middle-left and
middle-right plots, respectively, also demonstrate agreement between
the 5FS and the 4FS, with the central value of 4FS lying at the end of
the 5FS uncertainty band. The 5FS uncertainty band is relatively
large, reaching $\pm 40\%$ for transverse momenta larger than
$100~\textrm{GeV}$ and $50~\textrm{GeV}$ for the hardest and softest
$b$-jets, respectively. The uncertainty band for the 4FS prediction is
smaller, but only takes the renormalisation and factorisation scale
dependence into account, and is therefore a lower limit on the true
uncertainty.

In the lower-left plot, the $\Delta R$ separation between the two
leading $b$-jets is shown. For small $\Delta R_{bb}$, the 5FS
uncertainties are large, and the 4FS prediction lies within this
uncertainty band. For $\Delta R_{bb}\gtrsim 1.5$ the 5FS uncertainty
band shrinks to approximately ${}^{+10\%}_{-20\%}$, but the 4FS scale
dependence band increases to approximately ${}^{+25\%}_{-30\%}$. Due
to this increase of uncertainty in the 4FS calculations, both schemes
are in agreement for this observable.  Finally, in the lower-right
plot of Fig.~\ref{fig_2b}, the transverse momentum of the light jet is
shown. The ratio between the central values of the 4FS and the 5FS
predictions for this observable in the one-$b$-jet and two-$b$-jet
scenarios is very similar. On the other hand, the uncertainty band for
the 5FS is considerably larger in the two-$b$-jet scenario compared to
the one-$b$-jet scenario, resulting in a better agreement between
these two flavour schemes in the former, as compared to the latter.

\section{Conclusions}\label{conclusion}
In this study, we have presented a calculation for the \ttbb\ process
in the 5FS at NLO accuracy for the 13~TeV LHC. Our approach involved
computing the inclusive $t\bar{t}+\textrm{jets}$ process using the
FxFx merging prescription with up to 2 jets at NLO accuracy and
matching these predictions to the \pythia\ parton shower. We examined
the phase-space region containing at least one additional $b$-jet
alongside the top quark pair, as well as the region with at least two
additional $b$ jets. To improve the efficiency of selecting these jets
within the inclusive sample, we implemented a method to enhance the
probability of producing short-distance events with additional bottom
quarks in the \madgraph\ code, compensating for this enhancement by
reducing the weight of these events. This approach enabled us to
produce distributions in the one-$b$-jet and two-$b$-jet phase-space
regions with only modest statistical uncertainties, starting from just
5 million $t\bar{t}+\textrm{jets}$ short-distance events. Our results
demonstrate the viability of producing the \ttbb\ process in the 5FS
at NLO accuracy, yielding the most accurate predictions for this
process to date.

Compared to predictions at NLO+PS in the 4FS, we observed sizeable
differences with the 5FS. Notably, events in the 5FS exhibit higher
energy levels, with the transverse momentum of the top quark pair and,
to a somewhat lesser extent, the transverse momentum of the hardest
light jet displaying a harder spectrum in the 5FS. When taking the
uncertainty bands into account, for all observables considered the
bands either overlap or touch -- and care should be taken here that
the uncertainty estimated for the 5FS is more reliable than the one
for the 4FS, since the latter are expected to have significant
uncertainties stemming from the matching scheme which are not taken
into account.

Based on our findings, we advocate for the use of the 5FS in
predicting the \ttbb\ process at the LHC. The 5FS approach offers
enhanced accuracy at a modest efficiency cost, providing valuable
insights into the physics of top quark pair production in association
with bottom quarks. Ultimately, the improvements in efficiency
discussed in this work can also be applied to the ``fusion''
method~\cite{Hoche:2019ncc,Ferencz:2024pay}, allowing one to upgrade
that method also with the 5FS component to be computed at NLO accuracy
for up to 2 jets.

%
\section*{Acknowledgments}
RF is supported by the Swedish Research Council contract numbers
2016-05996 and 2020-04423. TM is supported by the German Research Foundation 
(DFG) under grant No.~AR 1321/1-1 and the Freiburg Scientific Society.

\appendix
\section{Top quark pair transverse momentum}
\begin{figure}[!t]
\centering
\includegraphics[width=0.45\textwidth]{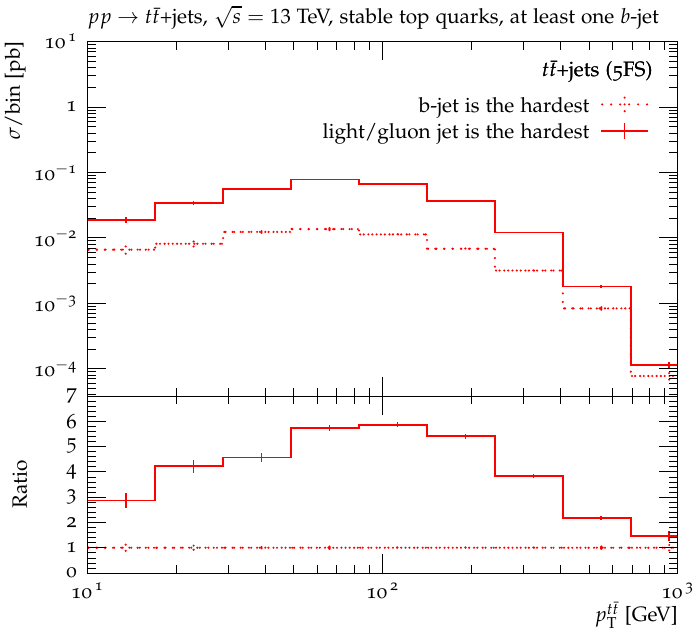}
\includegraphics[width=0.45\textwidth]{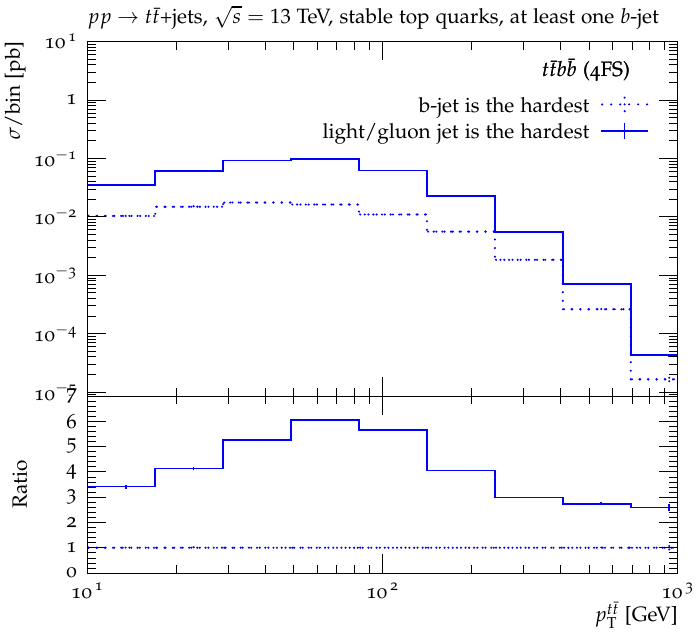}
\includegraphics[width=0.45\textwidth]{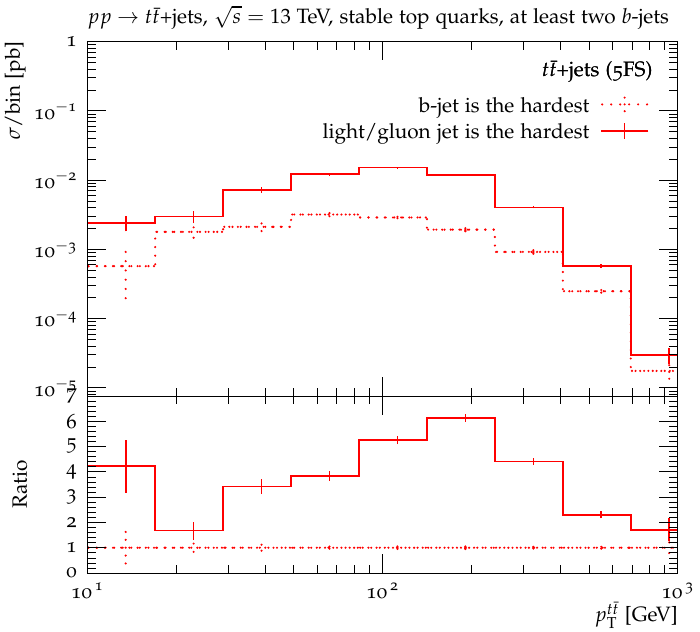}
\includegraphics[width=0.45\textwidth]{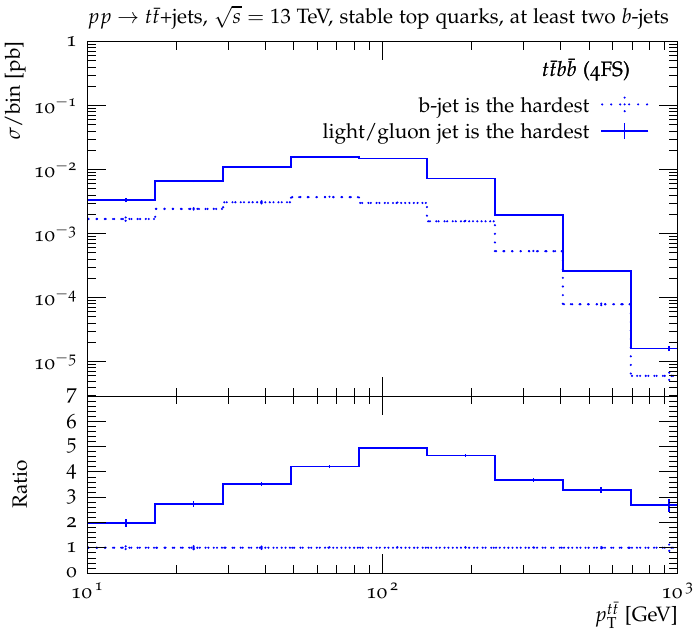}
\caption{Transverse momentum of the top quark pair in the one-$b$-jet
  (top row) and two-$b$-jet (bottom row) scenarios. On the left are
  the predictions in the 5FS and on the right in the 4FS. The
  distribution is split into the case where the hardest jet is a
  $b$-jet and the hardest jet is a light jet. }
\label{fig_3}
\end{figure}

The top quark transverse momentum, $p_T^{t\bar{t}}$, distribution in
the one-$b$-jet and two-$b$-jet scenarios show a notable difference in
shape in the 5FS and 4FS predictions, see figs.~\ref{fig_1b}
and~\ref{fig_2b}. In fig.~\ref{fig_3} we split this observable into
flavour of the hardest jet. The rationale behind this observable is
that at large transverse momentum, it is kinematically most-likely
that the transverse momentum of the top quark pair recoils against a
single hard jet. If the hardest jet is a light jet, then this jet is
described at NLO accuracy in the 5FS, while in the 4FS light jets are
described only at tree level or by the parton shower. As can be seen
in fig.~\ref{fig_3}, it is more likely that the hardest jet is light
jet than a $b$-flavoured jet for the whole range of $p_T^{t\bar{t}}$
considered, for both the 4FS and 5FS. For $p_T^{t\bar{t}}\gtrsim 100
\textrm{~GeV}$, there are significant differences between the 4FS and
5FS predictions: the 5FS predicts that more often the hardest jet is
non-$b$-flavoured for $p_T^{t\bar{t}}$’s up to about 500 GeV, after
which the ratio in the 4FS flattens out, while in the 5FS it keeps
going down. The latter is compatible with the distributions for the
transverse momenta of the hardest light and $b$-jets: at very high
$p_T$, the ratio of the 5FS over the 4FS predictions is larger for
$b$-jets than for light jets. This indeed shows that at least at
large-pT, the correlation between $p_T^{t\bar{t}}$ and
$p_T$(light-jet, hardest) (and to a lesser extend $p_T$($b$-jet,
hardest)), together with the difference between the 5FS and 4FS in
fraction of events for which the hardest jet is $b$-flavour as
compared to light, is the reason for the large difference between the
4FS and 5FS predictions.

%
\bibliography{ttbb_5F} 
\bibliographystyle{JHEP}

\end{document}